\documentclass[%
reprint,
superscriptaddress,
 amsmath,amssymb,
 aps,
pra,
]{revtex4-2}

\usepackage{graphicx}% Include figure files
\usepackage{dcolumn}% Align table columns on decimal point
\usepackage{bm,placeins}% bold math
\usepackage{amsmath,amssymb}
\usepackage{subfigure}
\usepackage{mdframed}
\usepackage{hyperref}
\usepackage{appendix}
\newcommand{\ket}[1]{\ensuremath{\left|#1\right>}}

\begin{document}

\title{Multi-mode Time-delay Interferometer for Free-space Quantum Communication}

\author{Clinton Cahall}
\altaffiliation{Corresponding author: clinton.cahall@duke.edu}
\affiliation{Department of Electrical and Computer Engineering, Duke University, Durham, NC 27708, USA}
\author{Nurul T. Islam}
\affiliation{Department of Physics, The Ohio State University, Columbus, OH 43210, USA}
\author{Daniel J. Gauthier}
\affiliation{Department of Physics, The Ohio State University, Columbus, OH 43210, USA}
\author{Jungsang Kim}
\affiliation{Department of Electrical and Computer Engineering, Duke University, Durham, NC 27708, USA}
\affiliation{IonQ, Inc., College Park, MD 20740, USA}

\date{\today}% It is always \today, today,
             %  but any date may be explicitly specified

\begin{abstract}

Quantum communication schemes such as quantum key distribution (QKD) and superdense teleportation provide unique opportunities to communicate information securely. Increasingly, optical communication is being extended to free-space channels, but atmospheric turbulence in free-space channels requires optical receivers and measurement infrastructure to support many spatial modes. Here we present a multi-mode, Michelson-type time-delay interferometer using a field-widened design for the measurement of phase-encoded states in free-space communication schemes. The interferometer is constructed using glass beam paths to provide thermal stability, a field-widened angular tolerance, and a compact footprint. The performance of the interferometer is highlighted by measured visibilities of $99.02\pm0.05\,\%$, and $98.38\pm0.01\,\%$ for single- and multi-mode inputs, respectively. Additionally, high quality multi-mode interference is demonstrated for arbitrary spatial mode structures and for temperature changes of $\pm1.0\,^{\circ}$C. The interferometer has a measured optical path-length drift of $130\,$nm$/\,^{\circ}$C near room temperature. With this setup, we demonstrate the measurement of a two-peaked, multi-mode, single-photon state used in time-phase QKD with a visibility of $97.37\pm 0.01\,\%$.

\end{abstract}

\maketitle

\section{Introduction}

Optical communication protocols using time-bin or pulse-position modulation encoding are a good fit for free-space communication systems. Temporal states having bin widths on the order of a few hundred pico-seconds are robust against errors caused by dispersion from the atmosphere~\cite{kral2005optical}, while achieving high data rates. One significant challenge of establishing a free-space optical link is the distortion of the spatial mode of the beam at the receiver due to atmospheric turbulence. A free-space quantum communication scheme, such as quantum key distribution (QKD) using a time-phase encoding~\cite{brougham2016information,islam2017provably}, requires a time-delay interferometer designed to support many spatial modes to perform measurements in the phase-basis.

Previous work has demonstrated classical and quantum interference with multi-mode time-delay interferometers that leverage various methods to enable a high quality interference~\cite{vallone2016interference,zeitler2016super,jin2018demonstration}. An optical beam with many spatial modes could potentially suffer degraded interference due to each mode having a slightly different wavevector $k$, and therefore a different propagation path length through the interferometer. Robust multi-mode interferometers are designed to minimize the path length difference for each mode to allow a complete interference for an arbitrary spatial mode.

One technique used to reduce the mode-dependent path length uses a 4\,-$f$\,-\,imaging system in each arm to relay the mode during propagation and therefore remove the $k$-vector dependence. This approach has demonstrated fringe visibility as high as $93\%$~\cite{zeitler2016super}. Another promising design for enabling high-quality multi-mode interference is a field-widened Michelson-type interferometer without relay optics~\cite{shepherd1985wamdii,gault1985optimization}. One such design has reported $91\,\%$ interference visibility and $85\,\%$ entanglement visibility~\cite{jin2018demonstration}. In this article we discuss the design, construction, and characterization of a multi-mode time-delay interferometer operating at $1550\,$nm. The design presented here uses a Michelson-type layout without relay optics to achieve robust multi-mode interference. Our design features different glass materials in each beam path to achieve improved spatial mode and thermal performance in a compact layout. We demonstrate the application of the interferometer to a single-photon phase-state measurement used in a time-phase QKD scheme.

\section{Multi-Mode Delay Interferometer Design}

\begin{figure}[th]
\begin{center}
\includegraphics[width=1.0\linewidth]{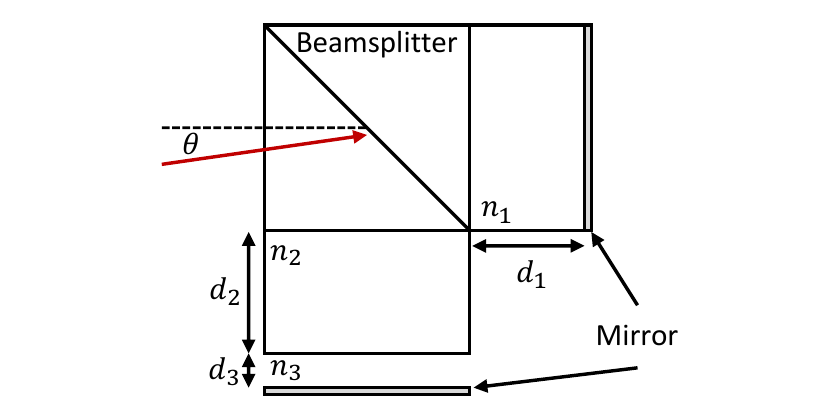}
\caption[Simplified schematic of a Michelson-type interferometer.]{Simplified schematic of a Michelson-type interferometer. This setup includes a small air gap at the end of the long arm ($d_{2}$).}
\label{fig:Int}
\end{center}
\end{figure}

A simplified interferometer layout is shown in Fig.~\ref{fig:Int}~\cite{gault1985optimization}. A ray incident from the left at an angle $\theta$ is split by a non-polarizing beamsplitter and each beam is directed to one of two pathways. One path has a short propagation length relative to the other. The short arm is made of material with index $n_{1}$ and thickness $d_{1}$, and the long arm is made of material with index $n_{2}$ and thickness $d_{2}$. Optionally, an air gap of thickness $d_{3}$ can be introduced between the end of the material in the long arm and the reflector. The two beams are reflected at the end of each arm and recombine at the beamsplitter.

The optical path length difference $\Delta$ between the two arms of the interferometer as a function of incident angle $\theta$ is given by~\cite[]{shepherd1985wamdii}

\begin{equation}\label{eq:OPD_1}
\begin{split}
\Delta = 2[n_{3}d_{3}(1-\sin^{2}\theta/n^{2}_{3})^{1/2}\\
+ n_{2}d_{2}(1-\sin^{2}\theta/n^{2}_{2})^{1/2}\\
- n_{1}d_{1}(1-\sin^{2}\theta/n^{2}_{1})^{1/2}],
\end{split}
\end{equation}

\noindent
where $n_{3}=1$ but is kept in the equation for completeness. Typically, a Michelson interferometer is designed for the chief ray to enter at normal incidence and small angular deviations from the chief ray are analyzed. In the design presented here, the input ray must enter at a finite angle $\theta_{0}$ to spatially separate each output beam from the input beam. Access to both outputs are required for the time-phase QKD protocol, discussed later. Unlike the analysis performed in~\cite{shepherd1985wamdii}, where Eq.~\ref{eq:OPD_1} is expanded about a normal incidence, we perform the expansion of Eq.~\ref{eq:OPD_1} about a finite input angle $\theta_{0}$. The first three terms of this expansion are given by

\begin{equation}\label{eq:OPD_2}
\begin{split}
\Delta = 2\left(d_{3}\beta_{3} +  d_{2}\beta_{2} -  d_{1}\beta_{1}\right)\\
 - 2\Phi(\sin\theta - \Phi)\left(\frac{d_{3}}{\beta_{3}} +  \frac{d_{2}}{\beta_{2}} -  \frac{d_{1}}{\beta_{1}}\right)\\
- (\sin\theta - \Phi)^{2}\left(\frac{n^{2}_{3}d_{3}}{\beta^{3}_{3}} +  \frac{n^{2}_{2}d_{2}}{\beta^{3}_{2}} -  \frac{n^{2}_{1}d_{1}}{\beta^{3}_{1}}\right),
\end{split}
\end{equation}

\noindent
where $\Phi = \sin\theta_{0}$ is the input angle and $\beta_{i}^{2} = n_{i}^{2} - \Phi^{2}$ is the effective index of refraction. The goal of our analysis is to satisfy three conditions that are derived from Eq.~\ref{eq:OPD_2}. A full mathematical derivation can be found in Ref.~\cite{cahall2019enabling}.

% A full derivation of the mathematical results is given in Appendix A.

\section{Design Requirements}

The interferometer is designed to achieve the following; (1) an optical path-length difference (OPD) of 200\,ps (5\,GHz free spectral range), (2) satisfy the field-widened condition (discussed more below) for a large angular bandwidth, and (3) be thermally compensated. A successful design will feature high visibility near unity, long-term temperature stability over a modest temperature range ($\pm0.2^{\circ}$\,C), and have phase-tuning ability over a complete interference fringe.

%\subsection{Optical Path-Length Difference}

The OPD of the chief ray is determined by setting $\sin\theta = \Phi$, leaving only the first term of Eq.~\ref{eq:OPD_2}. The path length difference $\Delta$ is in units of length, and therefore must be converted to a time delay $\Delta_{t}$ by dividing it by the speed of light in vacuum $c$ and given by 
\begin{equation}\label{eq:TimeDelay}
\Delta_{t} = 2(d_{3}\beta_{3} +  d_{2}\beta_{2} -  d_{1}\beta_{1})/c.
\end{equation}

%\subsection{Field-Widening}
\noindent
A wide angular bandwidth is necessary for a multi-mode interferometer design because a multi-mode beam has a range of wavevectors and each will take a slightly different path through the interferometer. To achieve high visibility, the change in OPD as a function of input angle must be minimized. Field-widening is achieved by proper material choices and device geometry that make the second term in Eq.~\ref{eq:OPD_2} vanish. The field-widening condition is
\begin{equation}\label{eq:FieldWiden}
 \frac{d_{3}}{\beta_{3}} + \frac{d_{2}}{\beta_{2}} - \frac{d_{1}}{\beta_{1}} = 0.
 \end{equation}

\noindent 
The result of satisfying Eq.~\ref{eq:FieldWiden} is that the OPD in Eq.~\ref{eq:OPD_2} becomes first-order insensitive to deviations from the input angle of the chief ray~\cite{gault1985optimization,shepherd1985wamdii}.

%\subsection{Thermal Compensation}

Lastly, thermal compensation provides long-term stability in an imperfectly controlled environment. Temperature fluctuations cause OPD changes due to expansion and contraction of the physical lengths of the glass material, as well as temperature-dependent changes in the index of refraction~\cite{shepherd1985wamdii}. The thermal drift is characterized by taking the partial derivative of Eq.~\ref{eq:OPD_2} with respect to temperature $T$, as

\begin{equation}\label{eq:ThermalDrift}
\begin{split}
\frac{\partial\Delta}{\partial T} = 2d_{2}\left[\beta_{2}\alpha_{2} + \frac{n_{2}}{\beta_{2}}\left(\frac{\partial n_{2}}{\partial T}\right)\right]\\
- 2d_{1}\left[\beta_{1}\alpha_{1} + \frac{n_{1}}{\beta_{1}}\left(\frac{\partial n_{1}}{\partial T}\right)\right],
\end{split}
\end{equation}

\noindent
where $\alpha_{i}=(1/d_{i})(\partial d_{i}/\partial T)$ is the coefficient of thermal expansion (CTE). This analysis assumes that the index and length change of the air gap with temperature is negligible and therefore those terms vanish when taking the derivative. The interferometer is made a-thermal by setting $\partial\Delta/\partial T = 0$, imposing a constraint on the path lengths $d_{1}$ and $d_{2}$, given by

\begin{equation}\label{eq:Athermal}
\left[\beta_{1}\alpha_{1} + \frac{n_{1}}{\beta_{1}}\left(\frac{dn_{1}}{dT}\right)\right]d_{1} = \left[\beta_{2}\alpha_{2} + \frac{n_{2}}{\beta_{2}}\left(\frac{dn_{2}}{dT}\right)\right]d_{2}.
\end{equation}

\section{Performance Simulation}

We conducted a survey of optical glass with the goal of identifying two glass types to construct the interferometer (See Appendix~\ref{appendix:material}) . The basic guidelines we used for making the glass choices are: (1) availability, (2) a large difference in index of refraction between the two materials, and (3) small relative changes in refractive index and length with temperature. A large difference in the index of refraction will enable smaller physical lengths to satisfy a given time-delay.

The finalized design uses N-SF66 and N-BK10 (Schott AG) for the long arm and short arm respectively. At the design wavelength of 1550\,nm, N-BK10 has an index of refraction of $n_{1}=1.4823$, $\partial n_{1}/\partial T = 1.4154\times10^{-6}\,\text{K}^{-1}$, and a CTE of $\alpha_{1}=5.8\times10^{-6}\,\text{K}^{-1}$. N-SF66 has an index of refraction of $n_{2}=1.8660$, $\partial n_{1}/\partial T = -1.5575\times10^{-6}\,\text{K}^{-1}$, and a CTE of $\alpha_{1}=5.9\times10^{-6}\,\text{K}^{-1}$. The index and change in index for each material are calculated at $\lambda=1550\,$nm using the Sellmeier equation and its derivative with respect to temperature. The glass rods are mounted to a substrate of Kovar nickel alloy. The CTE of Kovar is $\alpha_{\text{Kovar}}=5.86\times10^{-6}\,\text{K}^{-1}$ and is closely matched to the glasses used in our design. A good match between the glass and substrate, rather than using an ultra-low expansion (ULE) substrate with a much smaller CTE, is advantageous because matching CTE values between the glass and substrate will minimize the stress on the glass during temperature changes, which can cause birefringence and refraction.

The lengths $d_{1}$ and $d_{2}$ are optimized for an input angle $\theta_{0}=3^{\circ}$ and satisfy the OPD, field-widening, and a-thermal conditions. However, satisfying these three conditions simultaneously leads to large physical dimensions of the interferometer ($>50$\,mm). The negative consequence of large glass pathways are: (1) large glass path lengths increase the thermal mass to be stabilized and can increase the settling time for phase tuning: (2) as the path length increases so does the clear aperture required for a beam input at a fixed input angle; and (3) longer path-lengths will be more sensitive to overlap misalignment at the beamsplitter. Misalignment is especially sensitive for multi-mode interference because the spatial frequency of the mode can easily be on the order of the alignment tolerance, whereas in single-mode interference, the mode size is large compared to the alignment tolerance.

The goal of the design presented here is to minimize the physical dimensions of the glass beam paths. Minimizing the glass length will allow for a more robust and manufacturable setup. To meet this goal, we relax the thermal requirements of the design so that the a-thermal condition is satisfied only approximately, while the OPD and field-widening conditions are satisfied exactly. We find a unique solution that satisfies Eq.~\ref{eq:TimeDelay} and Eq.~\ref{eq:FieldWiden} by eliminating the air gap $d_{3}$. The OPD of the chief ray and field-widening condition for an interferometer without an air gap $d_{3}$ are given by

\begin{equation}\label{eq:TimeDelay2}
\Delta_{t} = 2(d_{2}\beta_{2} -  d_{1}\beta_{1})/c,
\end{equation}
and
\begin{equation}\label{eq:FieldWiden2}
\frac{d_{2}}{\beta_{2}} - \frac{d_{1}}{\beta_{1}} = 0,
\end{equation}
respectively. The a-thermal condition remains unchanged when removing the air gap.

\begin{figure}[th]
\begin{center}
\includegraphics[width=1.0\linewidth]{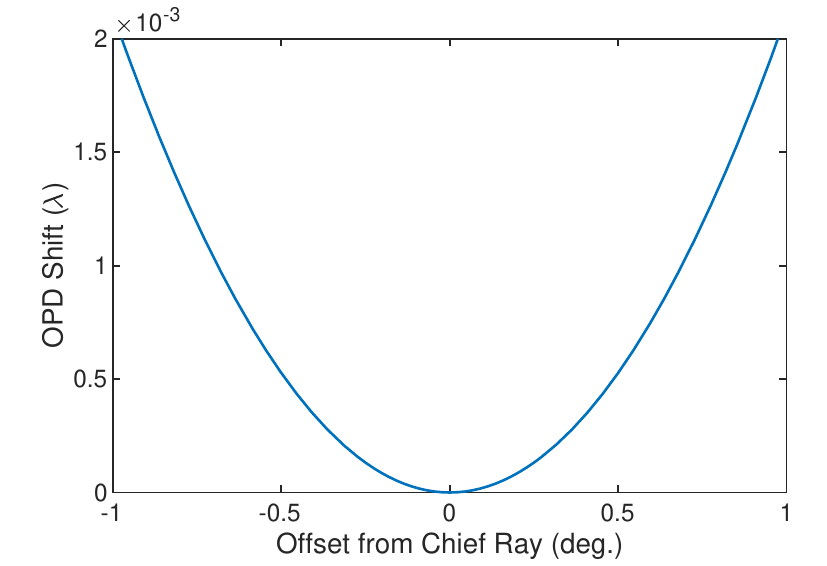}
\caption[Angular bandwidth about the chief ray of the multi-mode interferometer.]{Simulated change in the optical path difference as a function of deviations of the input about the chief ray of the multi-mode interferometer.}
\label{fig:OPDangle}
\end{center}
\end{figure}

The absence of an air gap leaves two equations and two free parameters, $d_{1}$ and $d_{2}$. The path lengths that satisfy Eq.~\ref{eq:TimeDelay2} and Eq.~\ref{eq:FieldWiden2} are $d_{1} = 34.5\,$mm and $d_{2} = 43.5\,$mm. At an input angle of $\theta_{0}=3^{\circ}$ and a beam diameter $\sim2\,$mm (discussed in more detail below), the required clear aperture is $\sim10\,$mm. A clear aperture of this size allows the square cross-section of the glass rods to be $12.7\,\text{mm}\times12.7\,\text{mm}$, and leads to manageable total thermal mass. Additionally, a 10\,mm clear aperture is $<80\,\%$ of a 12.7\,mm surface, which allows each surface to be polished easily with standard glass processing capabilities.

\begin{figure}[th]
\begin{center}
\includegraphics[width=1.0\linewidth]{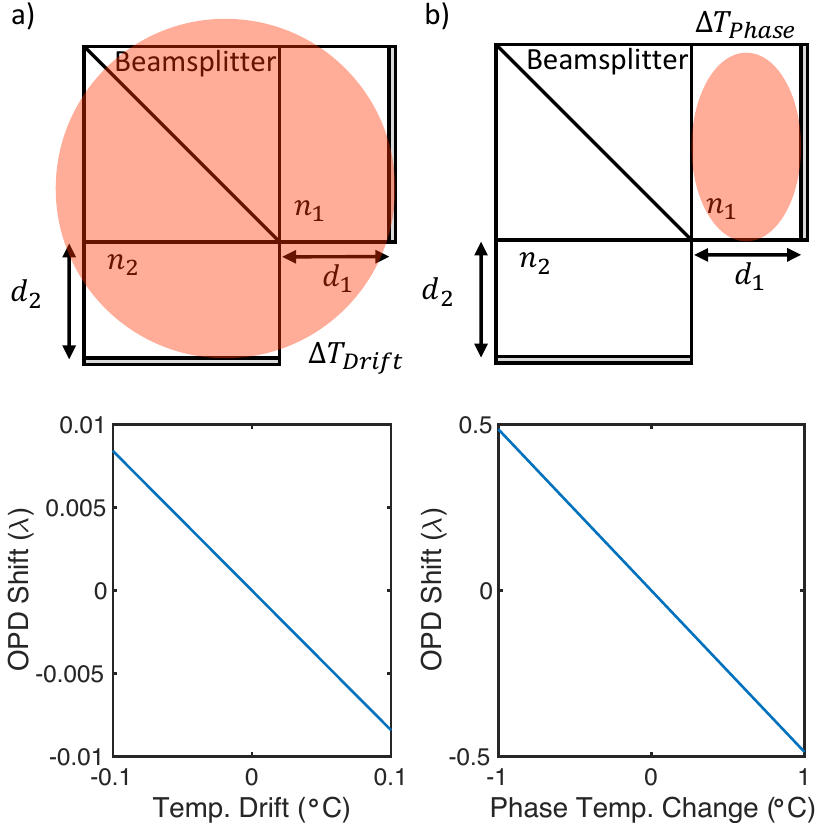}
\caption[Simulated thermal performance of the multi-mode interferometer.]{Simulated thermal performance of the multi-mode interferometer. (a) Thermal stability for small temperature drifts of the complete system. (b) Interference phase tuning via control over the temperature of the short arm $d_{1}$.}
\label{fig:OPDdrift}
\end{center}
\end{figure}

The angular bandwidth of the interferometer is characterized by plotting the OPD in Eq.~\ref{eq:OPD_2} for a range of input angles $\theta$ around the chief ray $\Phi = \sin3^{\circ}$, shown in Fig.~\ref{fig:OPDangle}. Recall that satisfying the field-widening condition given in Eq.~\ref{eq:FieldWiden2} implies that the first term in the angular dependence of the OPD (Eq.~\ref{eq:OPD_2}) is proportional to $\sin^{2}\theta$.

The thermal stability of the setup is estimated by calculating Eq.~\ref{eq:ThermalDrift} for small temperature changes $\Delta T$, shown in Fig.~\ref{fig:OPDdrift}(a). The simulation shows that thermal drift in the OPD of $\leq\lambda/50$ is achievable for modest temperature fluctuations of $\pm0.1^{\,\circ}$C. The phase of the interference is adjusted by controlling the temperature in one of the glass arms, the short arm for example, to $\pm1^{\,\circ}$C. Temperature control over this range will result in a change in the OPD from $-\lambda/2$ to $\lambda/2$, allowing for a tuning range of $0$ to $2\pi$ for the phase $\phi$, as shown in Fig.~\ref{fig:OPDdrift}(b). Thermal tuning is preferred over mechanical tuning because mechanical tuning techniques such as piezo-driven mirrors can be expensive, require an extra control and feedback system, and have large thermal drift.

\section{Interference Characterization}

\begin{figure}[th]
\begin{center}
\includegraphics[width=1.0\linewidth]{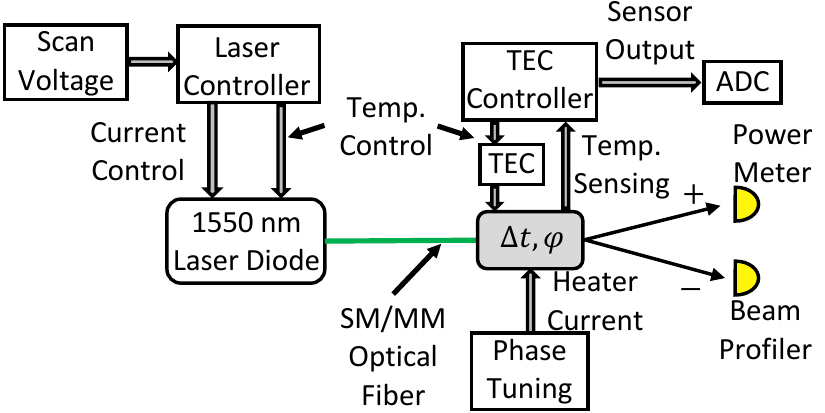}
\caption{Schematic of the experimental setup used to characterize the interference quality and thermal performance of the interferometer (gray box). SM: single-mode, MM: multi-mode, TEC: thermoelectric cooler, ADC: analog-to-digital converter.}
\label{fig:System}
\end{center}
\end{figure}

The performance of the interferometer is verified using optical beams with single-mode (SM) and multi-mode (MM) intensity profiles. Atmospheric turbulence is emulated by mode mixing in a perturbed multi-mode fiber~\cite[]{ursin2007entanglement,jin2018demonstration}. The experimental setup, shown schematically in Fig.~\ref{fig:System}, consists of a continuous-wave laser diode (Fitel F0L15DCWC) operating at 1550\,nm with a laser controller (Arroyo 6310 Combo Source) that controls the diode current and the temperature of the package. The wavelength of the output for diodes of this type is a function of the diode current and temperature. The controller has an analog voltage input for manual adjustment of the current supplied to the laser diode. A controllable voltage source is used as the scan voltage input for diode current sweeps. Adhesive heater tape (Omega KHLVA-101/5-P) is used on the short arm to provide phase tuning of the interferometer. The light from the laser is directed to the interferometer via a single or multi-mode optical fiber. The output of the interferometer is coupled in to an optical fiber where the output power is measured with an optical power meter (Newport Optical Power Meter 1803C with Model 818-IG sensor head), or the intensity profile of the beam can be viewed with a beam profiling camera (Ophir-Spricon SP907-1550).

\begin{figure}[!ht]
\begin{center}
\includegraphics[width=1.0\linewidth]{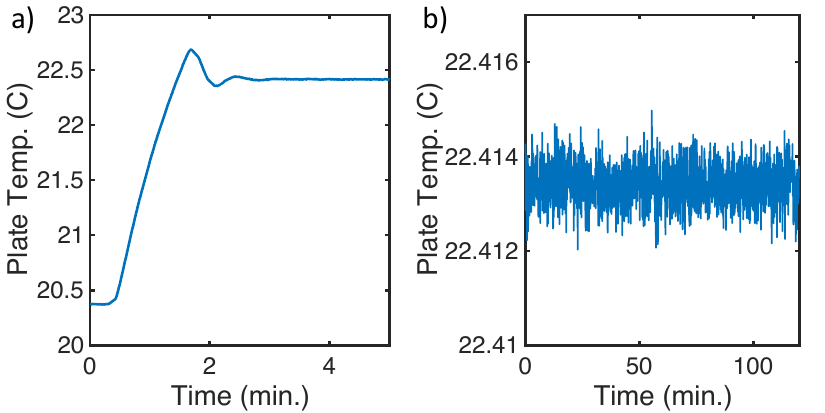}
\caption{Temperature control of the interferometer bottom plate. (a) Settling time of the plate temperature after a step change of $2^{\,\circ}$C. The target temperature is stable after $\sim3$ minutes. (b) Long-term temperature stability at a fixed set point. The temperature is controlled with an accuracy of $\pm0.0015^{\,\circ}$C.}
\label{fig:TempSet}
\end{center}
\end{figure}

In a practical setting, the temperature of the interferometer is meant to be passively controlled by modest isolation from the surrounding lab environment. For the purposes of thermal characterization, the experimental setup includes precise temperature control with thermo-electric cooling elements (TEC, Laird 64975-502) and a TEC controller (Thorlabs TED200C) with proportional-integral-derivative (PID) feedback control. Three TEC elements are placed between the bottom of the interferometer base plate and a large thermal bath, which is the optical table in this case. The temperature of the base plate is sensed via high-accuracy thermistors (Vishay NTCALUG02A103F161) placed at various locations on the plate. A thermistor near the center of the plate is used for the TEC feedback control. The analog output from the TEC controller is digitized with an analog-to-digital converter (ADC, National Instruments NI-9239) and recorded. A representative example of the settling time and temperature stability of the interferometer is shown in Fig.~\ref{fig:TempSet}.

\begin{figure}[th]
\begin{center}
\includegraphics[width=1.0\linewidth]{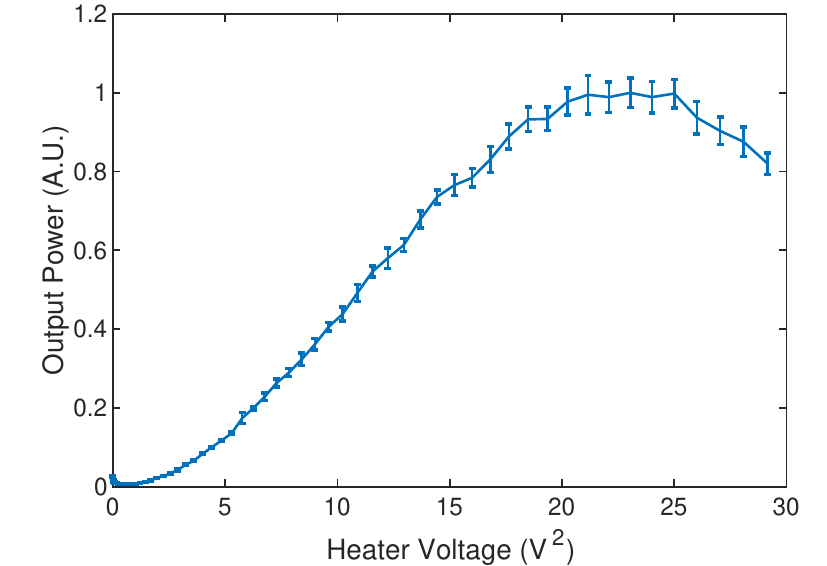}
\caption{Interference phase tuning with short arm temperature control. The output from the interferometer is plotted as a function of the square of the voltage supplied to the heater element on the short arm showing a tuning range over a full interference fringe.}
\label{fig:PhaseTune}
\end{center}
\end{figure}

\subsection{Fringe Visibility}

The phase of the interference fringe is scanned by heating one of glass arms and the result is shown in Fig.~\ref{fig:PhaseTune}. To maintain repeatability and execute measurements on faster timescales than thermal settling allows, interference fringes are recorded by scanning the input wavelength rather than scanning the phase of the interferometer. The wavelength of the laser diode is adjusted with an analog voltage input to the laser controller, which controls the current supplied to the diode. As the laser current is scanned, the wavelength changes but so does the total output power.  The output power as a function of the laser current is first calibrated by measuring the power directly from the laser. The change in optical power as a function of diode current is compensated for by dividing the interference fringe measurements by the calibration measurement. The power output of the interferometer is recorded as a function of the input scan voltage value. The quality of the interference is quantified by the visibility $\mathcal{V}$, given by

\begin{equation}\label{eq:Vis}
\mathcal{V}=\frac{\mathcal{P}_{+}-\mathcal{P}_{-}}{\mathcal{P}_{+}+\mathcal{P}_{-}}
\end{equation}

\noindent
where $\mathcal{P}_{+}$ is the maximum and $\mathcal{P}_{-}$ is the minimum value of the interference fringe.

\begin{figure}[th]
\begin{center}
\includegraphics[width=1.0\linewidth]{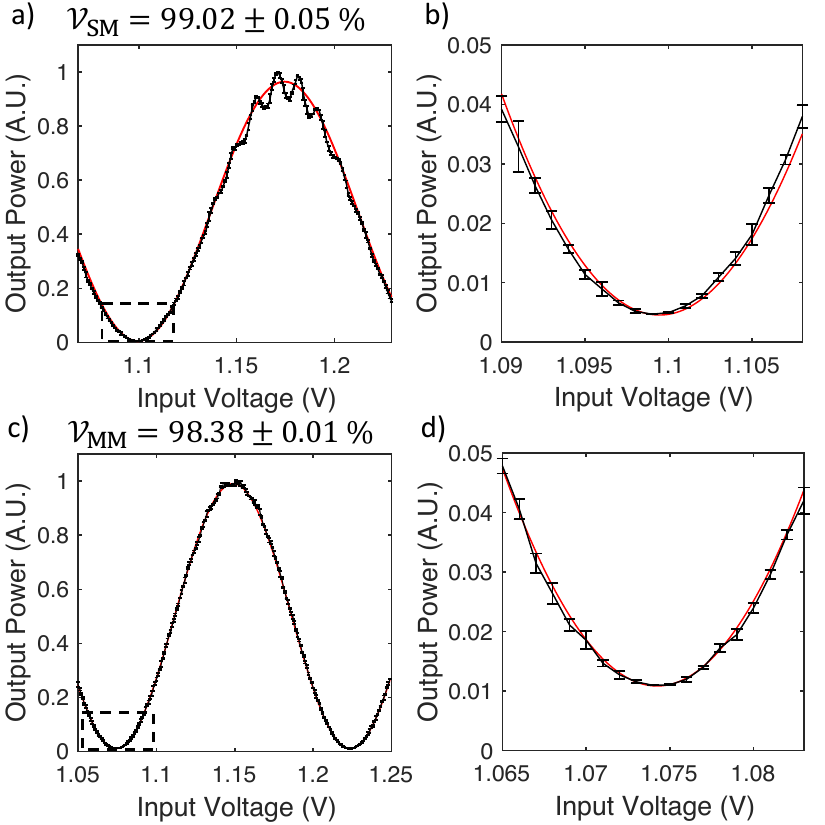}
\caption[Interference fringes for SM and MM ($50\,\mu$m core fiber) input beams.]{Interference fringes for SM and MM ($50\,\mu$m core fiber) input beams as a function of the input voltage to scan the laser diode current. Each data point (black) is the average of 10 measurements and error bars indicate the standard deviation of those 10 measurements. Red lines show a fit to the data using the function given by Eq.~\ref{eq:Pout}. (a) SM interference fringe. The visibility of the interference is $\mathcal{V}_{\mathrm{SM}}=99.02\pm0.05\,\%$. The high order fringe structure in the constructive interference could be due to reflections from multiple surfaces within the interferometer. (b) Zoomed view of the SM fringe minimum. (c) MM interference fringe. The visibility of the interference is $\mathcal{V}_{\mathrm{MM}}=98.38\pm0.01\,\%$. (d) Zoomed view of the MM fringe minimum.}
\label{fig:Fringes}
\end{center}
\end{figure}

The power $P_{\mathrm{out},\pm}$ measured at the $+/-$ port of the interferometer as a function of small changes of the wavelength $\delta\lambda\ll\Delta$, when the phase of the interference is set to $\phi$, is given by~\cite[]{islam2017robust}

\begin{equation}\label{eq:Pout}
\frac{P_{\mathrm{out},\pm}}{P_{0}}=\frac{a}{2}\left[1\mp\sin\left(k \delta\lambda + \phi\right)\right] + b,
\end{equation}

\noindent
where $P_{0}$ is the input optical power, $a\in\left\{0,1\right\}$ is the insertion loss of the interferometer, $b$ accounts for imperfect destructive interference in the minimum of the fringe, and $k=2\pi/\lambda$ is the wavenumber of the light. The output power is recorded as a function of the laser diode scan voltage, which serves as the wavelength adjustment parameter. The fringe measurement is fit with a curve given by Eq.~\ref{eq:Pout} with $a$, $b$, and $\phi$ as fitting parameters. The visibility of the normalized fringe is extracted from this fit where

\begin{equation}\label{eq:Vis2}
\mathcal{V}=\frac{a}{a + 2b}
\end{equation}

The measurement results from a single-mode input fiber (Corning SMF-28-Ultra) and multi-mode input fiber ($50\,\mu$m core, 0.1\,NA, Coastal Connections M-FUFU-50sL) are shown in Fig.~\ref{fig:Fringes}. The interferometer demonstrates single-mode interference visibility $\mathcal{V}_{\mathrm{SM}}=99.02\pm0.05\,\%$ and multi-mode interference visibility $\mathcal{V}_{\mathrm{MM}}=98.38\pm0.01\,\%$ extracted from the fitted functions. The single-mode results are close to those reported by high-performing commercial interferometers such as those used in a recent QKD demonstration~\cite{islam2017robust,islam2017provably}. Additionally, the multi-mode results presented here are significantly better than those recently reported for multi-mode interferometers~\cite{zeitler2016super,jin2018demonstration}.

\subsection{Spatial Mode Performance}

A robust multi-mode interferometer is required to maintain high visibility independent of the input spatial mode structure. The multi-mode results presented above are optimized and performed on a single beam profile. To investigate the robustness of the performance, we characterize the visibility of the interference fringe as the spatial profile is changed.  The visibility is maximized for the first spatial profile by tuning the beam alignment such that there is good overlap at the beamsplitter. Then, interference fringes are recorded for subsequent spatial profiles after the mode has been changed. The difference between spatial mode profiles is quantified by the two dimensional Pearson correlation coefficient $R$ defined as

\begin{equation}\label{eq:2dcorr1}
R = \frac{\sum_{i=1}^{N_{x}}\sum_{j=1}^{N_{y}}\left[I_{ij}\left(t_{1}\right)-\mu\left(t_{1}\right)\right]\left[I_{ij}\left(t_{2}\right)-\mu\left(t_{2}\right)\right]}{\sigma(t_{1})\sigma(t_{2})},\\[1.0em]
\end{equation}

\noindent
where $I_{ij}(t_{n})$ is the optical intensity collected at time $t_{n}$ (a frame), indexed by a discrete array of positions $(i,j)$ that correspond to the pixels of the camera, $\mu\left(t_{n}\right)$ is the mean intensity, and $\sigma(t_{n})$ is the standard deviation of the intensity distribution. The correlation coefficient $R$ has a value between $+1$ and $-1$, where $+1 (-1)$ indicates perfect positive (negative) correlation, and values close to $0$ indicate no correlation.

\begin{figure}[th]
\begin{center}
\includegraphics[width=1.0\linewidth]{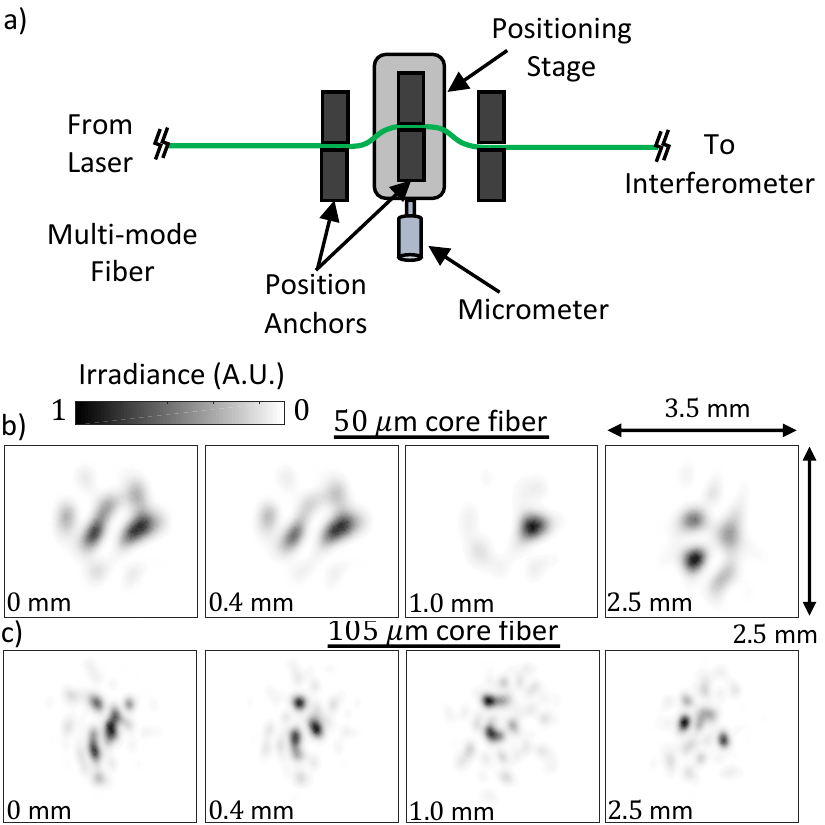}
\caption{Spatial mode characterization from a $50\,\mu$m and $105\,\mu$m core fiber. (a) Schematic showing the setup used to adjust the mode profile of the beam. The spatial mode is changed by inducing bends in the fiber controlled by a positioning stage and a micrometer. (b) Spatial mode examples for different micrometer settings from a $50\,\mu$m core fiber. (c) Spatial mode examples for different micrometer settings from a $105\,\mu$m core fiber.}
\label{fig:MMode}
\end{center}
\end{figure}

\begin{figure}[tb]
\begin{center}
\includegraphics[width=1.0\linewidth]{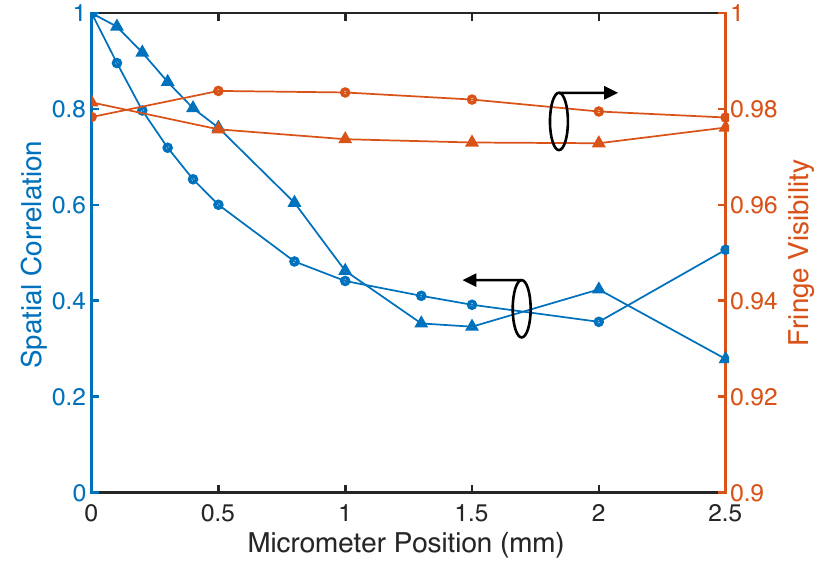}
\caption{Multi-mode interference as a function of the input spatial mode. Correlation coefficient between the first frame and subsequent frames as a function of the micrometer setting for $50\,\mu$m (circles) and $105\,\mu$m (triangles) core fibers is shown against the left vertical axis. Also plotted is the visibility of the interference fringe over the same micrometer range for each fiber, shown in the right vertical axis. Error in the fringe visibility is $\leq5\times10^{-4}$.}
\label{fig:VisCorr}
\end{center}
\end{figure}

Optical intensity profiles are collected by the beam profiling camera and the correlation coefficient is processed in MATLAB. The spatial mode of the beam is changed by introducing a controlled bend in the optical fiber with a micrometer and positioning stage. A schematic depicting the setup to induce changes in the mode is shown in Fig.~\ref{fig:MMode}(a). The fiber is anchored at three different points with the middle point on the positioning stage. As the position of the stage is adjusted, the fiber is bent at two different points that leads to the change in the mode profile. Special care is taken to ensure that all other points in the fiber are kept stationary throughout the experiment. When the fiber is kept secure apart from the micrometer adjustment, the spatial modes can be reproduced as a function of the micrometer position.

The profile of the beam is measured at the output of the fiber (without collimation) using the beam profiler. Examples of the intensity patterns from $50\,\mu$m and $105\,\mu$m core (Thorlabs FG105LVA) fibers for different micrometer positions are shown in Fig.~\ref{fig:MMode}(b) and (c). Note that the experiment is carried out by first recording the mode profiles with the beam profiler at each micrometer position, then performing the interference measurements at each micrometer position. Between the beam profiler measurements and the interference measurements, the output of the fiber must be moved from the beam profiler to the interferometer input. The movement of the fiber will unavoidably change the actual spatial mode that travels through the interferometer from those in the recorded images in Fig.~\ref{fig:MMode}(b) and (c). However, the intensity profile and the \emph{correlation} between profiles at different micrometer positions recorded by the beam profiler are representative of the actual mode passing through the interferometer. Therefore, the mode analysis using the beam profiler adequately represents the spatial mode at the input of the interferometer.

The correlation coefficient and interference visibility as a function of micrometer position are shown in Fig.~\ref{fig:VisCorr}. The correlation between the first measurement and subsequent measurements decays to a value $<0.5$ as the spatial mode becomes more distinguishable. As the spatial mode is changed, the interferometer maintains excellent fringe visibility for each measurement. The quality of the interference using a $50\,\mu$m core fiber has an average value $\mathcal{V}_{\mathrm{avg}}=98.05\pm0.01\%$ and a maximum value $\mathcal{V}_{\mathrm{max}}=98.38\pm0.01\%$. Additionally, the quality of the interference using a $105\,\mu$m core fiber has an average value $\mathcal{V}_{\mathrm{avg}}=97.55\pm0.04\%$ and a maximum value $\mathcal{V}_{\mathrm{max}}=98.14\pm0.02\%$. The quality of these results indicate an interferometer design that is robust for arbitrary input spatial modes.

\subsection{Thermal Performance}

\begin{figure}[thb]
\begin{center}
\includegraphics[width=1.0\linewidth]{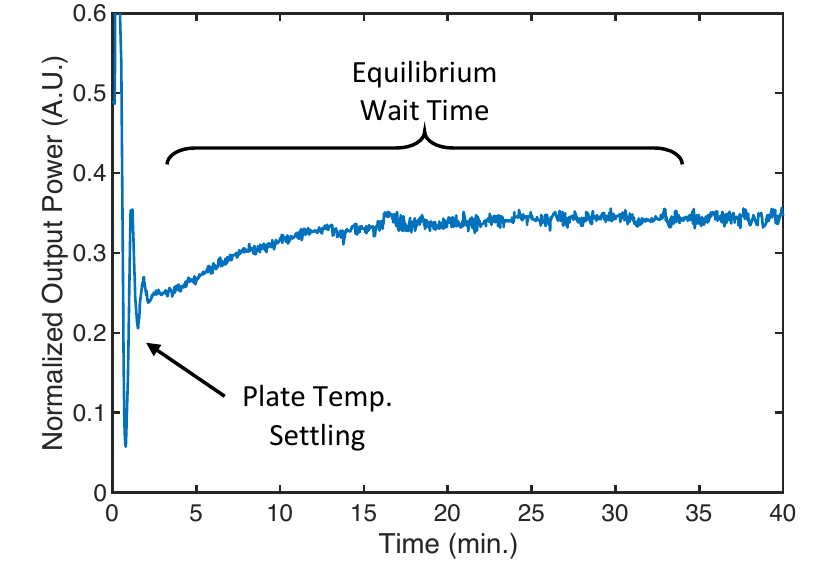}\\[1.1em]
\caption{Example measurement of the output power from the interferometer while the system reaches thermal equilibrium. This particular set of data shows a temperature change from $21.0^{\,\circ}$C to $21.5^{\,\circ}$C}
\label{fig:EqTime}
\end{center}
\end{figure}

The thermal performance of the interferometer is characterized by measuring the visibility of the interference fringe and the path length shift as a function of temperature. The temperature of the bottom plate is set and maintained by the TEC controller that supplies current to the TEC elements in contact with the substrate of the interferometer. When the setpoint is changed, the system is allowed to come to thermal equilibrium over a period of $\sim30$ minutes. As thermal equilibrium is reached, the path length will drift as a function of time, causing a change in the measured output power. Therefore, when a new setpoint is investigated, the output of the interferometer is monitored as equilibrium is reached. We infer that the interferometer has come to thermal equilibrium once the output power is stabilized, shown in Fig.~\ref{fig:EqTime}. During setpoint changes, the phase of the interference is set such that it lies near the point of the steepest slope to have the maximum sensitivity in determining the equilibrium.

\begin{figure}[thb]
\begin{center}
\includegraphics[width=1.0\linewidth]{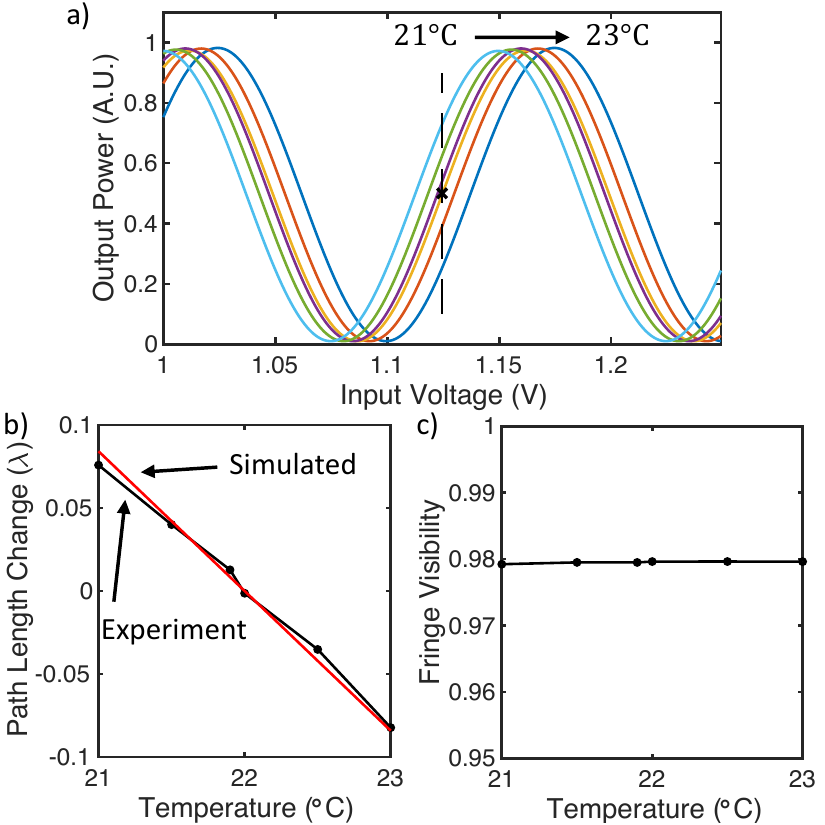}
\caption{Multi-mode interference as a function of temperature. (a) Fitted functions to the measured interference fringe at each temperature setting. The temperature settings for the curves moving from left to right are $21.0, 21.5, 21.9, 22.0, 22.5,$ and $23.0\,^{\circ}$C. The phase of the interference at $22.0\,^{\circ}$C is set at data point marked with an `X.' The path length shift is calculated using points where the fringe intersects the dotted line. (b) Experimentally measured (black) and theoretically predicted (red) path length change as a function of temperature. (c) Interference visibility as a function of temperature.}
\label{fig:TempScan}
\end{center}
\end{figure}

Once the system has come to thermal equilibrium, a full interference fringe is scanned and fit with the function given by Eq.~\ref{eq:Pout} in the same manner as discussed above. A fringe is recorded at each temperature setting and plotted as a function of the scanning voltage applied to the diode controller, corresponding to the change in the laser wavelength. The results of the curve fitting are shown in Fig.~\ref{fig:TempScan}. Path length drift brought on by the temperature change causes the fringe to move horizontally. While operating at a fixed value of $\phi$ (shown by the `X' on the yellow curve in the figure) a horizontal shift of the fringe manifests in a change in the power measured at the output port of the interferometer. The change in output power describes change in path length.

The path length change and visibility are measured for a temperature range of $\pm1.0\,^{\circ}$C centered at $22.0\,^{\circ}$C. The results of the measurements are shown in Fig.~\ref{fig:TempScan}(b) and (c). A high fringe visibility of $\sim98\,\%$ is maintained across the entire temperature range. The experimentally determined path-length shift as a function of temperature is 130\,nm/$\,^{\circ}$C. This value matches well with the theoretically predicted performance. Therefore, stability on the order of $\lambda/50$ can be achieved with moderate temperature control in a lab environment on the order of $0.25\,^{\circ}$C.

\section{Application to Time-Phase QKD}

Experimental realizations of QKD systems are often hindered by non-idealalities in the equipment used to generate and measure the photons. Rate limitations are dominated by detector saturation and dead-time, and so it is advantageous to increase the channel capacity by encoding bits in a high dimensional Hilbert space. We use a time-phase encoding~\cite{brougham2016information} because it can support a high dimensional encoding. An encoding scheme having $d$-dimensions is often referred to as a qudit protocol, where $d=2$ is the well-known qubit protocol. Qudits can transmit up to $\log_{2}d$ bits per photon, whereas qubit schemes, such as BB84~\cite{bennett1984quantum}, are limited to a single bit per photon. Additionally, high-dimensional schemes can tolerate a higher bit-error-rate, which enable longer communication distances~\cite{scarani2009security}. Recently, a high-dimensional QKD scheme using a $d=4$ time-phase encoding has achieved a record high maximum secret-key-rate (SKR) of 26.2\,Mbit/s, and an SKR of 1.07\,Mbit/s at an equivalent fiber distance of 83\,km~\cite{islam2017provably}.

A time-phase QKD protocol consist of two mutually unbiased bases that are referred to as the temporal (or time) basis and the phase (or frequency) basis. Data is encoded in the temporal basis and transmitted from Alice to Bob, and the phase basis is used to check for the presence of an eavesdropper, Eve. In general, $d-1$ cascaded interferometers are needed to perform a complete measurement of the phase basis~\cite{brougham2013security}. A full description of such a protocol, including a demonstration utilizing a high-dimensional encoding, can be found in Ref.~\cite{islam2017provably}. Here, we use a $d=2$ encoding to demonstrate the performance of our multi-mode interferometer.

\begin{figure}[t]
\begin{center}
\includegraphics[width=1.0\linewidth]{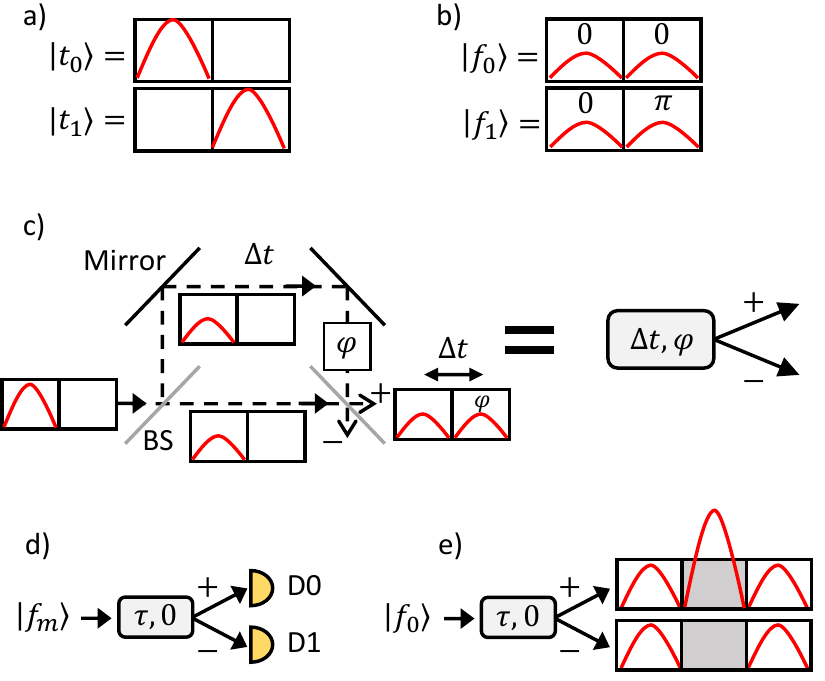}
\caption{Time-phase protocol with $d=2$. (a) Temporal and (b) phase basis states. (c) Time-delay interferometer and it's action on a single wavepacket. (d) Interferometer setup for a $d=2$ phase basis measurement. (e) Interference patterns for a measurement on the $\ket{f_{0}}$ state.}
\label{fig:2dTimePhase}
\end{center}
\end{figure}

The states for a $d=2$ time-phase protocol are shown in Fig.~\ref{fig:2dTimePhase}. Temporal states are constructed by intensity modulation of a continuous-wave (CW) laser source to create short duration wavepackets. The states are highly attenuated to produce weak coherent states with a mean photon number per pulse of $\mu\approx1$. Phase states are constructed using intensity modulators to create the superposition of time states and an additional phase modulator to adjust the phase of the wavepackets in each time bin. The phase states are attenuated so that the probability amplitudes of the entire frame sum to unity. It is important to use a modulation scheme with a high extinction ratio to prevent errors caused by background photons leaking through the modulators. 

A measurement in the temporal basis is performed with a single-photon detector, where the electrical signal generated from a detection event is recorded with a high resolution time-tagger. The single-photon detector and processing electronics, including amplifiers and time-tagging electronics, must have low timing uncertainty (detection jitter) to localize the time-of-arrival of the photon with high accuracy~\cite{marsili2013detecting,cahall2018scalable}. The arrival time of the single-photon detection event is compared to the shared global clock to declare the result of the measurement.

Phase-state measurements require a time-delay interferometer, shown schematically in Fig.~\ref{fig:2dTimePhase}(c). A phase state entering the interferometer is split by a 50/50 beamsplitter with half of the wavepacket traveling through a long path and half through a short path. The difference in path lengths is chosen such that the time-delay is equal to $\tau$, or one time bin width. Therefore, the wavepackets recombining at the second beamsplitter are shifted by one time bin relative to one another. An additional phase adjustment $\phi$ is available to tune the phase of the interference fringe, where $\phi=0$ in the $d=2$ scheme.

A phase state that passes through the interferometer will result in constructive or destructive interference in the shaded time-bin shown in Fig.~\ref{fig:2dTimePhase}(e). There is a one-to-one mapping between the phase-state input and the detector in which constructive interference is observed. Therefore, constructive interference is detected at a unique detector, with destructive interference at all other detectors. In the $d=2$ protocol, a photon detection event in the shaded time bin of detector $D0$ ($D1$) indicates a measurement outcome of $\ket{f_{0}}$ ($\ket{f_{1}}$). The lobes in the interference pattern lying outside the shaded time bin are considered losses and do not contribute to the outcome of the measurement. The phase-basis measurement succeeds at a rate of $1/d$, or $50\%$ for the $d=2$ scheme. The security of the channel is assessed at the end of all communication by calculating the visibility of the constructive and destructive interference fringes. The visibility of the interference is given by

\begin{equation}\label{eq:Vis3}
\mathcal{V}=\frac{\mathcal{P}_{0}-\mathcal{P}_{1}}{\mathcal{P}_{0}+\mathcal{P}_{1}}
\end{equation}

\noindent
where $\mathcal{P}_{0}$ ($\mathcal{P}_{1}$) is the summation of counts recorded by detector $D0$ ($D1$) in the shaded time bin. The visibility of interference has a direct impact on the length of the secret key that is distilled from the raw key. For the $d=2$ time-phase scheme discussed here, the quantum bit-error-rate (QBER) of a phase-basis measurement is given by~\cite{scarani2009security}

\begin{equation}\label{eq:QBER}
QBER = \frac{1-\mathcal{V}}{2}.
\end{equation}

\noindent
As the visibility of the interference increases the QBER decreases and therefore a larger secret key can be distilled from the raw key.

\begin{figure}[t]
\begin{center}
\includegraphics[width=1.0\linewidth]{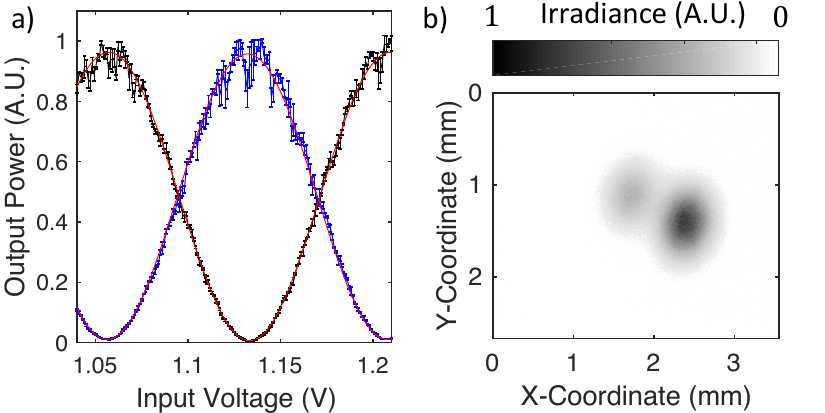}
\caption{Interference characterization and spatial mode for phase-state measurement. (a) Output from the plus (minus) port of the interferometer shown in blue (black) with Eq.~\ref{eq:Pout} fitted to each curve shown in red. The visibility between the plus and minus port with the phase set at the maximum of the plus port is $98.7\,\%$. (b) Example of the spatial mode under test measured at the output of a $25\,\mu$m core fiber.}
\label{fig:25um}
\end{center}
\end{figure}

Experimentally, the generation of the states shown in Fig.~\ref{fig:2dTimePhase}(a) and (b) are carried out in a similar manner to the method discussed in Ref.~\cite{islam2017robust}. A continuous-wave laser operating at $1550\,$nm  is intensity modulated with electro-optic modulators to create short duration ($\sim80$\,ps) wavepackets. The modulators are driven by a 10\,GHz transceiver from a field-programmable gate array (FPGA). The mean photon number for each wavepacket is controlled using a variable optical attenuator (VOA). The modulation and attenuation are performed in single-mode fiber, then the light is coupled to a multi-mode fiber with a $25\,\mu$m core (Thorlabs FG025LJA) to produce wavepackets with multiple spatial modes. Mode mixing is achieved by stressing the fiber in the same manner as shown in Fig.~\ref{fig:MMode}(a). A $25\,\mu$m core fiber is chosen over the $50\,\mu$m and $105\,\mu$m core fibers mentioned previously because of limitations in our single-photon detection system. The light is sent through the interferometer, where each output is collected with a $25\,\mu$m core fiber. The fibers are then coupled to our single-photon detection system.

Our detection system is comprised of amorphous superconducting nanowire single-photon detectors (SNSPDs) produced by Quantum Opus. SNSPDs are high speed and high efficiency single photon counting devices (see Ref.~\cite{dauler2014review} for a review) that are widely used in quantum communication~\cite{boroson2014overview,islam2017provably} and quantum optics experiments~\cite{shalm2015strong}. Though large area devices~\cite{wollman2017uv} and pixel arrays~\cite{verma2014four,allman2015near} designed to support free-space or multi-mode fiber coupling exist, our devices have an active area  with a diamter of $\sim15\,\mu$m and are designed to couple to a single-mode telecommunications fiber (eg. Corning SMF-28)~\cite{miller2011compact}. However, in this experiment, we couple the $25\,\mu$m core fibers from the output of the interferometer. There will be some amount of aperture loss at the detector due to the mismatch in the size of the detector and the fiber core. We choose to suffer the extra loss at the detector for the purpose of demonstrating multi-mode, single-photon interference. The fringes from the plus and minus port of the interferometer and an example of the spatial mode profile of the beam under test at the output of the fiber are shown in Fig.~\ref{fig:25um}.

\begin{figure}[thb]
\begin{center}
\includegraphics[width=1.0\linewidth]{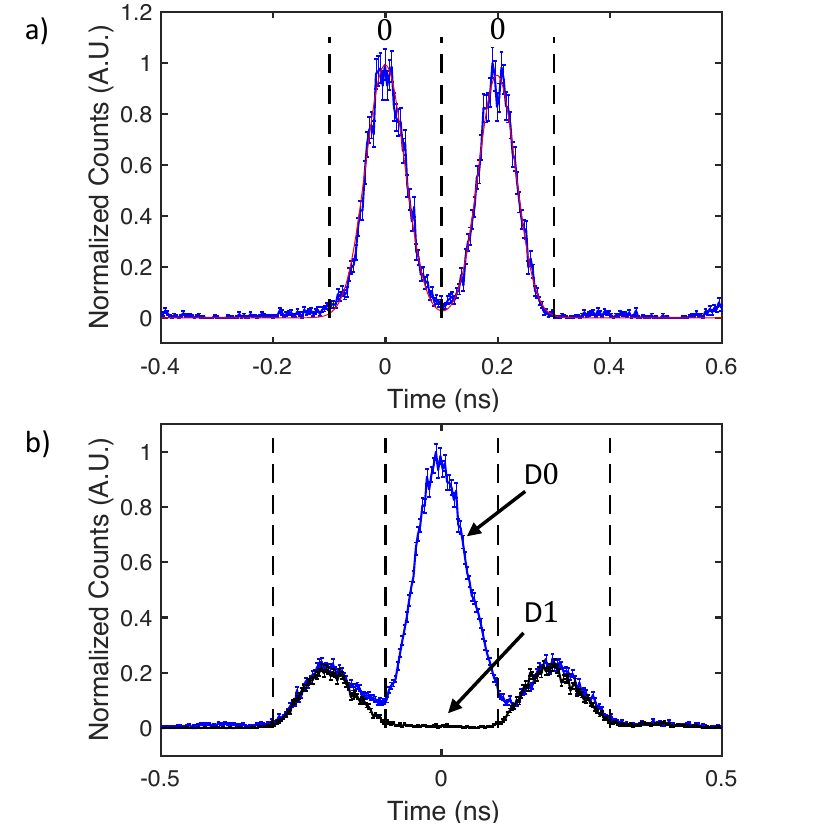}
\caption{Phase-state $\ket{f_{0}}$ preparation and measurement using a $25\,\mu$m core fiber. (a) Single-photon data (blue) with a sum of Gaussian functions fitted to the data (red) for the $\ket{f_{0}}$ state preparation. Zeros above the data denote the phase of each bin, indicating a $\ket{f_{0}}$ state. Time bin boundaries are shown with black dotted lines. (b) Measurement of the $\ket{f_{0}}$ state using the multi-mode interferometer showing the plus $(+)$ output port collected at detector D0 (blue) and minus $(-)$ output port collected at detector D1 (black). Time bin boundaries are shown with dotted lines. The resulting visibility between the bright and dark fringe in the middle time bin is $97.37\pm0.01\,\%$.}
\label{fig:Meas}
\end{center}
\end{figure}

The preparation of the $\ket{f_{0}}$ state is shown with single-photon detection events in Fig.~\ref{fig:Meas}(a). The error rate of the preparation, calculated from the overlap of each wavepacket in to the opposite time-bin, is $0.4\,\%$ and is largely due the limited bandwidth of the modulators used to create the wavepackets. The result of a measurement on the $\ket{f_{0}}$ state using the time-delay interferometer is shown in Fig.~\ref{fig:Meas}(b). As expected, constructive interference is observed in the middle time-bin of detector D0, with destructive interference observed in the middle time-bin of detector D1. Additionally, the heights of the lobes to the left and right of the middle time-bin are approximately $1/4$ the height of the constructive interference fringe, and match well with the predicted measurement outcome. The visibility between the two detectors in the middle time-bin is determined by summing all the counts falling in the time-bin window and using the formula given by Eq.~\ref{eq:Vis3}. The resulting visibility is $97.37\pm 0.01\,\%$. This value is within $1\,\%$ of the visibility reported for a similar phase-state measurement using commercial single-mode time-delay interferometers~\cite{islam2017robust}.

\section{Conclusion}

In this article, we detail the design methodology, performance simulations, construction and characterization of a time-delay interferometer with a 5\,GHz free-spectral-range (200\,ps time-delay). The interferometer is set-up in a Michelson-type layout with glass beam paths in each arm. A wide field-of-view and excellent thermal stability can be achieved with a proper material choice and optical design, while satisfying the target time-delay. The performance results are highlighted by a maximum single-mode visibility of $\mathcal{V}_{\mathrm{SM}}=99.02\pm0.05\,\%$, comparable to commercial single-mode products, and multi-mode visibility of $\mathcal{V}_{\mathrm{MM}}=98.38\pm0.01\,\%$, several percent higher than recent reports. Additionally, multi-mode visibility $\mathcal{V}_{\mathrm{MM}}\approx98\,\%$ is maintained independent of spatial mode profile for both $50\,\mu$m and $105\,\mu$m core fibers. High-quality interference is demonstrated over a modest temperature range of $\pm1^{\,\circ}$C where the path-length shift of the interferometer is 130\,nm/$^{\,\circ}$C, enabling good thermal stability with modest temperature control. 

Finally, we demonstrate the application of our interferometer to a time-phase QKD protocol, where the presence of an eavesdropper is quantified by the interference visibility between the two output ports of a time-delay interferometer. The high visibility of our interferometer enables phase-state measurements with a low QBER, allowing efficient extraction of a secret key. We demonstrate a phase-state measurement using weak coherent states with visibility of $97.37\pm 0.01\,\%$ for $1550\,$nm photons from a $25\,\mu$m core fiber.

Further improvements can be made to the design and construction of the interferometer to achieve better performance. 
In our estimation the largest contributing factor to the degradation of the multi-mode interference visibility is an error in the length of the glass rods during polishing. Each piece of glass is undersized by $400-600\,\mu$m. The implication of this error is that the actual dimensions of the glass no longer satisfy the field-widening condition and therefore result in degraded interference, especially for multi-mode beam profiles. We were able to compensate for this error by a coarse adjustment of the position of one of the reflecting mirrors. Using a more precise method for measuring the position of the mirror, we expect the quality of multi-mode visbility could be improved to over $99\,\%$.

\begin{acknowledgments}
This work is supported by the Office of Naval Research (ONR) (N00014-13-1-0627) and the National Aeronautics and Space Administration (NASA) (NNX13AP35A).
\end{acknowledgments}

\appendix

\section{Material Choice}
\label{appendix:material}

\begin{table}[th]

\begin{tabular}{|p{5em}||p{4em}|p{4em}|p{4em}|p{4em}|}
  \hline
 Glass Name & Index $n$ & CTE ($10^{-6}/K$) & $dn/dT$ ($10^{-6}/K$) & Mfr.\\
 \hline

S-FSL5 & 1.4732 & 9.0 & -1.1 & Ohara\\
N-BK10 & 1.4823 & 5.8 & 1.4 & Schott\\
N-BK7 & 1.5009 & 7.1 & 1.1 & Schott\\
S-NSL3 & 1.5025 & 9.0 & 0.3 & Ohara\\
S-BAL14 & 1.5512 & 8.0 & 1.3 & Ohara\\
S-TIH6 & 1.7628 & 8.9 & -0.4 & Ohara\\
N-SF6 & 1.7631 & 9.0 & -2.3 & Schott\\
S-NPH53 & 1.7997 & 7.4 & -0.8 & Ohara\\
SF57 & 1.8019 & 8.3 & 6.0 & Schott\\
S-NPH2 & 1.8614 & 6.7 & -0.2 & Ohara\\
L-LAH86 & 1.8615 & 6.1 & 4.1 & Ohara\\
N-SF66 & 1.8660 & 5.9 & -1.5 & Schott\\
S-LAH79 & 1.9552 & 6.0 & 7.0 & Ohara\\
N-LASF35 & 1.9747 & 8.5 & 2.7 & Schott\\
 \hline
\end{tabular}
\caption{Selection of glass options investigated for the construction of the interferometer.\label{table:mat}}
\end{table}

A survey of glass options was conducted and narrowed to the list of materials given in Table~\ref{table:mat}. The factors that led to the final decision of materials used are a combination of performance, availability, and cost.

\section{Interferometer Design and Construction}
\label{appendix:design}

\begin{figure}[th]
\begin{center}
\includegraphics[width=1.0\linewidth]{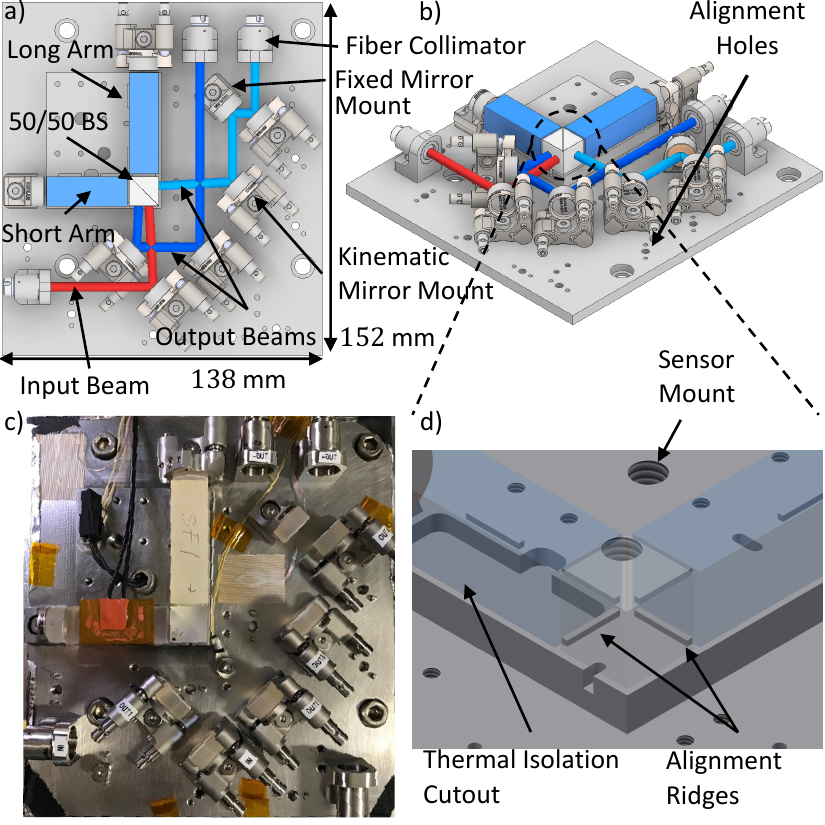}
\caption[CAD schematic of the interferometer with input and output beam steering and coupling optics.]{CAD schematic of the interferometer with input and output beam steering and coupling optics. Overhead CAD view (a) and perspective view (b) of the interferometer layout. (c) Photograph of the assembled interferometer. The heater tape is shown on the short arm glass and several temperature sensors are placed at various positions on the Kovar plate. (d) Detailed view of the beamsplitter and glass rod alignment features, and the substrate cutout for thermal isolation of the short arm.}
\label{fig:Construction}
\end{center}
\end{figure}

Construction of the interferometer is carried out with N-BK10 and N-SF66 glass rods measuring $12.7\,\text{mm}\times12.7\,\text{mm}\times34.5\,\text{mm}$ and $12.7\,\text{mm}\times12.7\,\text{mm}\times43.5\,\text{mm}$ respectively, where the long dimension is the optical direction. The end faces are optically polished and coated with an anti-reflective (AR) dielectric coating that is optimized for 1550\,nm. The glass rods are mounted to the Kovar nickel alloy substrate.

A computer-aided design (CAD) model of the complete interferometer system with input and output fiber coupling and beam steering optics is shown in Fig.~\ref{fig:Construction}. The input fiber is collimated with a fixed collimating package (f $=8\,$mm asphere, Thorlabs F240-1550) that is mounted in a custom-machined mount made of stainless steel.  The optical plane is defined by the center of the collimating lenses and the center of the mirrors, both of which are 12.7\,mm from the surface of the Kovar plate. The non-polarizing beamsplitter (Newport 05BC16NP.11) and glass rods are mounted on a platform to place the center of these pieces on the optical plane. The beamsplitter and glass rods are aligned on the substrate via 1\,mm-height, 1\,mm-width raised ridges.

\section{Deviation from Design}
\label{appendix:construction}
There are two important differences between the schematic in Fig.~\ref{fig:Int} and the simulated system discussed previously. The assembled system has 1\,mm gaps between the beamsplitter and glass rods, and 1\,mm gaps between the end of the glass rods and the reflecting mirrors in each arm. Air gaps are inserted to avoid the need for optical bonds at each interface to reduce complexity and increase the production quality of the assembly. Additionally, air gaps between the beamsplitter and glass rods allow each face to be AR coated and reduce the loss at these interfaces. Symmetric air gaps at the interfaces do not contribute to the OPD. Expansion and contraction of the substrate due to temperature changes will also be symmetric across each spacer gap and do not invalidate the mathematical analysis presented.

\bibliography{biblist}% Produces the bibliography via BibTeX.

\end{document}